\documentclass[11pt]{article}
\usepackage{amssymb}
\usepackage{graphics}
\usepackage{epsfig}
\usepackage{a4wide}
\usepackage{color}

\usepackage[utf8x]{inputenc}

\usepackage{amsmath}
\usepackage{amsthm}
\usepackage{amssymb}
\usepackage{graphicx}
\usepackage{epsfig}
\usepackage{verbatim}
\usepackage{slashed}
\usepackage{color}

\begin{document}
\title{ Enhancement of Loop induced 125GeV Higgs pair production through Large-Extra-Dimensions model at the LHC }
\author{ \textcolor{red}{}
Sun Hao$^{1}$\footnote{haosun@mail.ustc.edu.cn}, Zhou Ya-Jin$^2$\footnote{zhouyj@sdu.edu.cn} \\
{\small $^{1}$ School of Physics and Technology, University of Jinan, Jinan, Shandong 250022, P.R.China} \\
{\small $^{2}$ School of Physics, Shandong University, Jinan,
Shandong 250100, P.R.China} }
\date{}

\maketitle \vskip 15mm

\begin{abstract}
Based on the analysis of 5 $\text{fb}^{-1}$ of data at the LHC, the
ATLAS and CMS collaborations have presented evidence for a Higgs
boson with a mass in the 125 GeV range. We consider the 125 GeV
neutral Higgs pair production process in the context of
large-extra-dimensions(LED) model including the Kaluza-Klein(KK)
excited gravitons at the LHC. We take into account the LED effects
coming from gluon-gluon fusion and quark-antiquark collision
channels as well as their corresponding next-to-leading order(NLO)
QCD loop induced corrections. We analyse their impacts on both the
total cross section and some key distributions. Indeed,
$pp\rightarrow HH$ has the clear advantage of a lower standard
model(SM) background compare to process like $pp\rightarrow jj$,
though its SM prediction is very small, it is shown that the LED
model raises the cross section of Higgs pair production compare to
its SM prediction and enhance the transverse momentum($p_T^H$) and
invariant mass($M_{HH}$) distributions especially at high scales of
$p_T^H$ and $M_{HH}$. By including the NLO QCD loop corrections, the
scale dependence of total cross section can be reduced obviously.
Choose suitable decay modes like $HH\rightarrow
b\bar{b}\gamma\gamma$ or $HH\rightarrow b\bar{b}\mu^-\mu^+$ and some
simple cuts, we can strongly reduce the SM background but keep most
of the LED effects, leading Higgs pair production a promising
channel to search LED effects.

\hspace*{2cm}
\par
KeyWords: Higgs Pair Production, NLO QCD Corrections, Large Hadron Collider(LHC)
\end{abstract}

\vskip 3cm {\large\bf PACS: 11.10.Kk, 12.38.Bx, 14.80.Bn } 

\vfill \eject \baselineskip=0.32in
\renewcommand{\theequation}{\arabic{section}.\arabic{equation}}
\renewcommand{\thesection}{\Roman{section}.}
\newcommand{\nb}{\nonumber}
\newcommand{\Dir}{\kern -6.4pt\Big{/}}
\newcommand{\Dirin}{\kern -10.4pt\Big{/}\kern 4.4pt}
\newcommand{\DDir}{\kern -7.6pt\Big{/}}
\newcommand{\DGir}{\kern -6.0pt\Big{/}}
\makeatletter      
\@addtoreset{equation}{section}
\makeatother       
\vskip 5mm

\section{Introduction}
\par
The hierarchy problem of the standard model (SM) strongly suggests
new physics at TeV scale, and the idea that there exists extra
dimensions (ED) which first proposed by Arkani-Hamed, Dimopoulos,
and Dvali\cite{ADD} might provide a solution to this problem. They
proposed a scenario in which the SM field is constrained to the
common 3+1 space-time dimensions (``brane"), while gravity is free
to propagate throughout a larger multidimensional space $D=\delta+4$
(``bulk"). The picture of a massless graviton propagating in D
dimensions is equal to the picture that numerous massive
Kaluza-Klein (KK) gravitons propagate in 4 dimensions. The
fundamental Planck scale $M_S$ is related to the Plank mass scale
$M_{Pl}=G_N^{-1/2}=1.22\times10^{19}~{\rm GeV}$ according to the
formula $M^2_{Pl}=8\pi M^{\delta+2}_{S} R^\delta$ , where $R$ and
$\delta$ are the size and number of the extra dimensions,
respectively. If $R$ is large enough to make $M_S$ on the order of
the electroweak symmetry breaking scale ($\sim 1~ {\rm TeV}$), the
hierarchy problem will be naturally solved, so this extra dimension
model is called the large extra dimension model (LED) or the ADD
model. Postulating $M_S$ to be 1 TeV, we get $R\sim 10^{13}~{\rm
cm}$ for $\delta=1$, which is obviously ruled out since it would
modify Newton's law of gravity at solar-system distances; and we get
$R\sim 1~{\rm mm}$ for $\delta=2$, which is also ruled out by the
torsion-balance experiments\cite{Kapner:2006si}. When $\delta \geq
3$, where $R < 1~{\rm nm}$, it is possible to detect graviton signal
at high energy colliders.

\par
Both the ATLAS and CMS collaborations have reported a SM Higgs-like
excess at around $M_H=125$ GeV. If a SM-like Higgs particle is
discovered in this particular mass range, an important additional
test of the SM electroweak symmetry breaking sector is the
measurement of the Higgs self-interactions.
At hadron colliders, the pair production of Higgs bosons plays a
distinctive role in understanding the Higgs
mechanism\cite{HiggsMechanism}. As the triple self-coupling of Higgs
particles is involved in such production thus provide the
experimental reconstruction of the Higgs potential. Precise
measurement of this coupling could therefore give more insight on
the mechanism of electroweak symmetry breaking.
Compared to that of a single Higgs boson production, the
signal-to-background ratio could significantly improved. The
invariant mass scale of the single Higgs production is fixed by the
Higgs mass, of order only $\sim$ 125 GeV. Thus their detection
through heavy quark decay modes suffer from large QCD backgrounds.
Further more, the Higgs pair production can give various final states,
depending on the decay modes of the Higgs boson.
In the mass range of (120,130)GeV, the Higgs boson decay modes with
the largest branching fractions are $H\rightarrow b\bar{b}$ and $H\rightarrow W^+W^-$.
The most probable decay mode for a pair of Higgs bosons is $HH\rightarrow b\bar{b}b\bar{b}$.
However, this mode is challenging to search for due to the fact that it is
difficult to trigger on, and that it competes against the
QCD multi-jet backgrounds that possess overwhelmingly
large cross sections. In general, QCD backgrounds can
be suppressed with the existence of leptons and missing
energy. In the SM frame, the promising channels are
$pp\rightarrow HH\rightarrow b\bar{b}\tau^+\tau^-(b\bar{b}\mu^-\mu^+, b\bar{b}\gamma\gamma)$\cite{HH2bbrrbbmm}
and
$pp\rightarrow HH\rightarrow b\bar{b}W^+W^-$\cite{HH2bbtt} for $M_H\sim 125$ GeV.
More detail analysis can be found later.

\par
Another important distinctive feature of the Higgs pair production
at the LHC is that the effects of physics beyond the SM can
remarkably enhance the cross section with respect to that of the SM.
Phenomenological studies of Higgs pair production have thus been
performed in the context of the fourth generation
model\cite{sunhaopphh}, the littlest Higgs model\cite{wangleipphhLH}
and the Universal Extra Dimensions model\cite{pphhUED}. For the
large extra dimensional models, the tree level diagrams mediated by
the Kaluza-Klein gravitons lead to a large total cross section. Such
new theoretical approaches have drawn extensive attention in
ref\cite{pphhED} for a comparison between supersymmetry and LED
models. Exchange of virtual KK graviton or emission of a real KK
mode could give rise to interesting phenomenological signals at TeV
scale\cite{ADD:Gian,ADD:HanTao}. Virtual effects of KK modes could
lead to the enhancement of the cross section of pair productions in
processes, for example, di-lepton, di-gauge boson ($\gamma\gamma$,
$ZZ$, $W^+W^-$), dijet, $t\bar{t}$ pair\cite{ADDvirtualA,
ADDvirtualB, ADDvirtualC, ADDvirtualD, ADDvirtualE, ADDvirtualF}
etc. Some of these calculations have also been expanded into NLO QCD
loop level. A detailed calculation of Higgs pair production in LED
model has been performed recently\cite{pphhEDsunhao}, however, the
QCD loop induced calculation based on the LED context is still
missing.

\par
Experimentally, the CMS Collaboration has performed a lot of search
for LED on different final states at $\sqrt{s}=7$
TeV\cite{CMS:LED1,CMS:LED2,CMS:LED3}. By combining the diphoton,
dimuon and dielectron channels, lower limits are set on the
effective Planck scale in the range of 2.3$\sim$3.8 TeV at the
95$\%$ confidence level\cite{LEDbounds}. These limits are the most
restrictive bounds on virtual graviton exchange up to date. Based on
the analysis of 5 $fb^{-1}$ of data at the LHC, the
ATLAS\cite{SMHiggs125GeV_ATLAS} and CMS\cite{SMHiggs125GeV_CMS}
collaborations have presented evidence for a Higgs boson with a mass
in the 125 GeV range. We thus concentrate on the 125 GeV Higgs pair
production related to the latest measurement with the effects of the
LED models and find the characteristic distribution of it up to QCD
loop induced level.

\par
This paper is organized as follows: in section 2 we present a brief
introduction to the related theory. in section 3 we present the main
contribution to the process and section 3 is arranged to present the
numerical results of our studies. Finally we summarize the results
in the last section.

\section{Related theories}
\par
The LED model consists of the gravity sector and the SM sector. The
manifold which gravity propagates, is not the ordinary 4 dimensional
spacetime manifold, but R4$\times$M, where M is a $\delta$-torus
with radius $R$ and volume $V_{\delta}=(2\pi R)^{\delta}$ without
loss of physical significance. In our calculation we use the de
Donder gauge. The relevant Feynman rules involving spin-2 KK
graviton and the relevant vertices in the LED model can be found in
Ref\cite{ADD:Gian,ADD:HanTao,ADD:ppwwQCD:Mawengan}.

We denote the process as:
\begin{eqnarray}
pp\rightarrow ij\rightarrow G_{KK} \rightarrow
HH+X(ij=q\bar{q}={u\bar{u},d\bar{d},c\bar{c},s\bar{s},b\bar{b}},gg)
\end{eqnarray}
where $G_{KK}$ denotes the Kaluza-Klein(KK) gravitons. The couplings
between gravitons and SM particles are proportional to a constant
named gravitational coupling $\kappa \equiv \sqrt{16 \pi G_N}$,
which can be expressed in terms of the fundamental scale $M_S$ and
the size of the compactified space radius R by $ \kappa^2 R^{\delta}
= 8 \pi (4 \pi)^{\delta/2} \Gamma(\delta/2) M_S^{-(\delta+2)}$. In
practical experiments, the contributions of the different KK modes
have to be summed up, so the propagator is proportional to
$i/(s_{ij}-m^2_{\vec{n}})$, where $s_{ij}=(p_i+p_j)^2$ and
$m_{\vec{n}}$ is the mass of the KK state $\vec{n}$. When the
effects of all the KK states are taken together, the amplitude is
proportional to $\sum\limits_{\vec{n}}
\frac{i}{s_{ij}-m^2_{\vec{n}}+i\epsilon}=D(s)$. If $\delta \geq 2$
this summation is formally divergent as $m_{\vec{n}}$ becomes large.
We assume that the distribution has a ultraviolet cutoff at $\Lambda
\sim M_S$, where the underlying theory becomes manifest. Then $D(s)$
can be expressed as:
\begin{eqnarray}
 D(s) =\frac{1}{\kappa^2}
\frac{8\pi}{M_S^4}(\frac{\sqrt{s}}{M_s})^{\delta-2}[\pi + 2i
I(M_S/\sqrt{s})].
\end{eqnarray}
The imaginary part I($\Lambda/\sqrt{s}$) is from the summation over
the many non-resonant KK states and  its expression can be found in
Ref.\cite{ADD:HanTao}. Finally the KK graviton propagator after
summing over the KK states is:
\begin{eqnarray}
\label{prop} \tilde{G}^{\mu\nu\alpha\beta}_{KK}=D(s)
 \left(\eta_{\mu\alpha}\eta_{\nu\beta} + \eta_{\mu\beta}\eta_{\nu \alpha}
- \frac{2}{D-2}\eta_{\mu\nu}\eta_{\alpha\beta}\right)
\end{eqnarray}

\par
By adding all the Feynman rules in the LED model and the propagator
given above, we can get the amplitudes($\cal{M}$) for the SM and
virtual KK graviton exchange contributions as well as their
interferences. The cross section integral for hadron-hadron
collisions can be written as
\begin{equation}
\sigma=\sum_{i,j} \int^1_{\tau_0} \int^1_{\frac{\tau_0}{x_1}} dx_1
dx_2 G_{i/p_a}(x_1,\mu^2_f) G_{j/p_b}(x_2, \mu^2_f) \cdot \int
\frac{1}{\text{avgfac}} \frac{|{\cal M}_n ( \hat s = x_1 x_2 s
)|^2}{2 \hat s (2 \pi)^{3n-4}} d\Phi_n
\end{equation}
where $G_{i,j/p_a,p_b}(x,\mu^2)$ represent the gluon or (anti)quark
parton density functions, $p_a, p_b$ indicate (anti)proton, $\mu_f$
is the factorization scale which can be chosen equal the
renormalization scale $\mu_r$ when the loop calculation is included.
$\frac{1}{\text{avgfac}}$ is the times of spin-average factor,
color-average factor and identical particle factor. $|{\cal M}_n|^2$
presents the squared n-particle matrix element and divided by the
flux factor $[2 \hat s (2 \pi)^{3n-4}]$. $d\Phi_n$ denotes the
n-particle phase space differential. The quantity $\hat s = x_1 x_2
s$ is the effective center-of-mass(c.m.s.) energy, and the sum
$\sum_{i,j}$ runs in case of quark-antiquark incident partons over
all possible quark-antiquark combinations ($i,j=u,d,s,c,b,\bar u,
\bar d, \bar s, \bar c, \bar b$). In case of $gg$ initial state the
sum has only one term with $i,j=g$. Typically the latest new parton
distributions for collider physics CT10\cite{CT10} has been used in
our calculation.

\par
The n-body phase space differential $d\Phi_n$ and its integral
$\Phi_n$ depend only on $\hat s$ and particle masses $m_i$ due to
Lorentz invariance:
\begin{eqnarray} \nonumber
 \Phi_n(\hat s, m_1, m_2, ..., m_n) &=& \int d\Phi_n(\hat s, m_1, m_2,..., m_n) \\
&=&\int \delta^4((p_i+p_j)-\sum^{n}_{k=1}p_k) \prod^{n}_{k=1}d^4 p_k
\delta(p^2_k-m^2_k) \varTheta (p^0_{k})
\end{eqnarray}
with i and j denoting the incident particles and k running over all
outgoing particles(k = 1,...,n). In our calculation we use
BASES\cite{BASES} to do the phase space integration and
Kaleu\cite{Kaleu} to cross check with each other.

\section{Related Process}

\begin{figure}[hbtp]
\vspace{-5cm} \hspace*{-1.5cm} \centering
\includegraphics[scale=0.8]{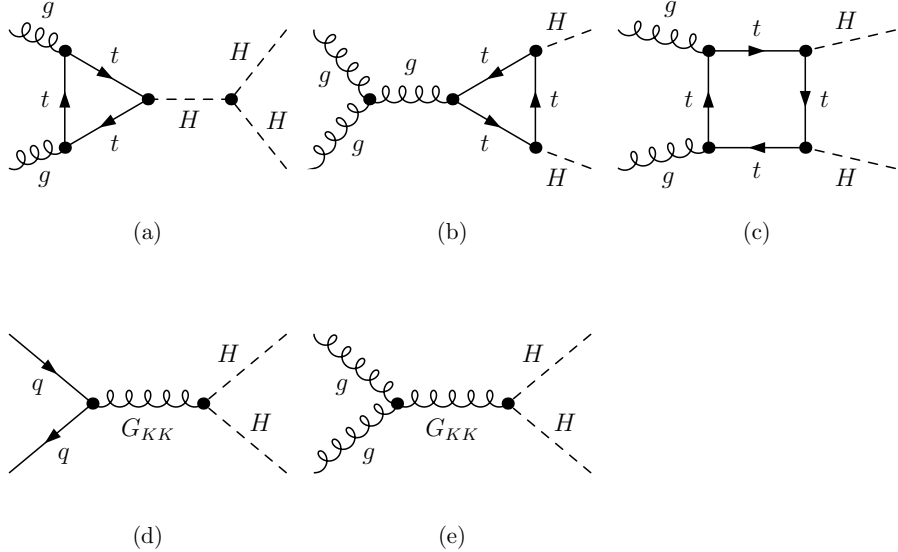}
\vspace{-10.5cm} \caption{\label{fig1} Part of Feynman diagrams for
$pp\rightarrow HH$ in the SM (a, b, c) and the tree level
quark-antiquark collision(d) and gluon-gluon fusion(e) diagrams in
LED model, where q represents u-,d-,c-,s-, and b-quark while
$G_{KK}$ represents spin-2 KK graviton.}
\end{figure}

\par
We separate the total contributions into two parts. One is the pure
SM effects and the other is the LED effects, simply denoted as
$\sigma^{SM}$ and $\sigma^{LED}$, respectively.

\par
In SM, there is gluon-gluon fusion channel contribute to SM
predictions through top-quark loops, see Fig. \ref{fig1}(a, b, c)
for more details. There is also b-quark contribution, but it is
small. Other diagrams include the change of the loop arrow in
Fig.\ref{fig1}(a, b, c) and cross change of the legs in Fig.
\ref{fig1}(c) which are similar thus not shown, totally ten Feynman
diagrams contribute. Contributions from the quark-antiquark
collision can be safely omitted in the light fermion mass limits
except b-b fusion through t-channels. However, it is only less of
0.5 percent of gluon-gluon fusion contribution\cite{sunhaopphh}, and
not considered here. Thus we can define this contribution as
$\sigma^{SM}=\sigma^{SM_{LOOP}}_{gg}$.

\par
Now let's see the second part: contributions to the LED effects
$\sigma^{LED}$. Several distinct contributions contribute to this
part, include both the tree level and QCD loop induced level. We
separate and highlight them into four different parts:

\begin{figure}[hbtp]
\vspace{-6cm} \hspace*{-1.4cm}
\includegraphics[scale=0.8]{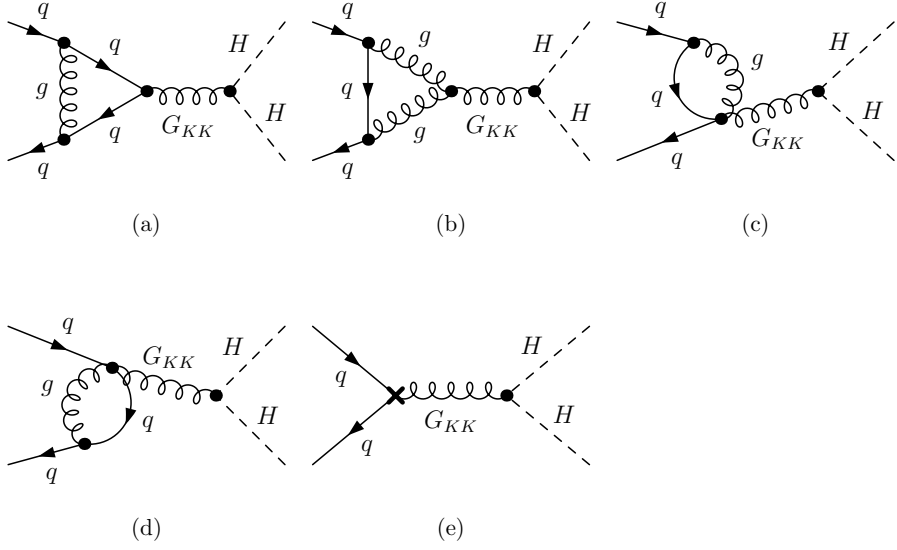}
\vspace{-11cm} \caption{\label{fig2} The QCD one-loop Feynman
diagrams for the partonic process $q\bar{q}\rightarrow G_{KK}
\rightarrow HH$ (a)-(d) and the counter term diagrams(e) correspond
to Fig.\ref{fig1}(d), where q represents u-,d-,c-,s-, b- and t-quark
while $G_{KK}$ represents spin-2 KK graviton. }
\end{figure}

\par
\begin{figure}[hbtp]
\vspace{-3.5cm} \hspace*{-1.0cm}
\includegraphics[scale=0.8]{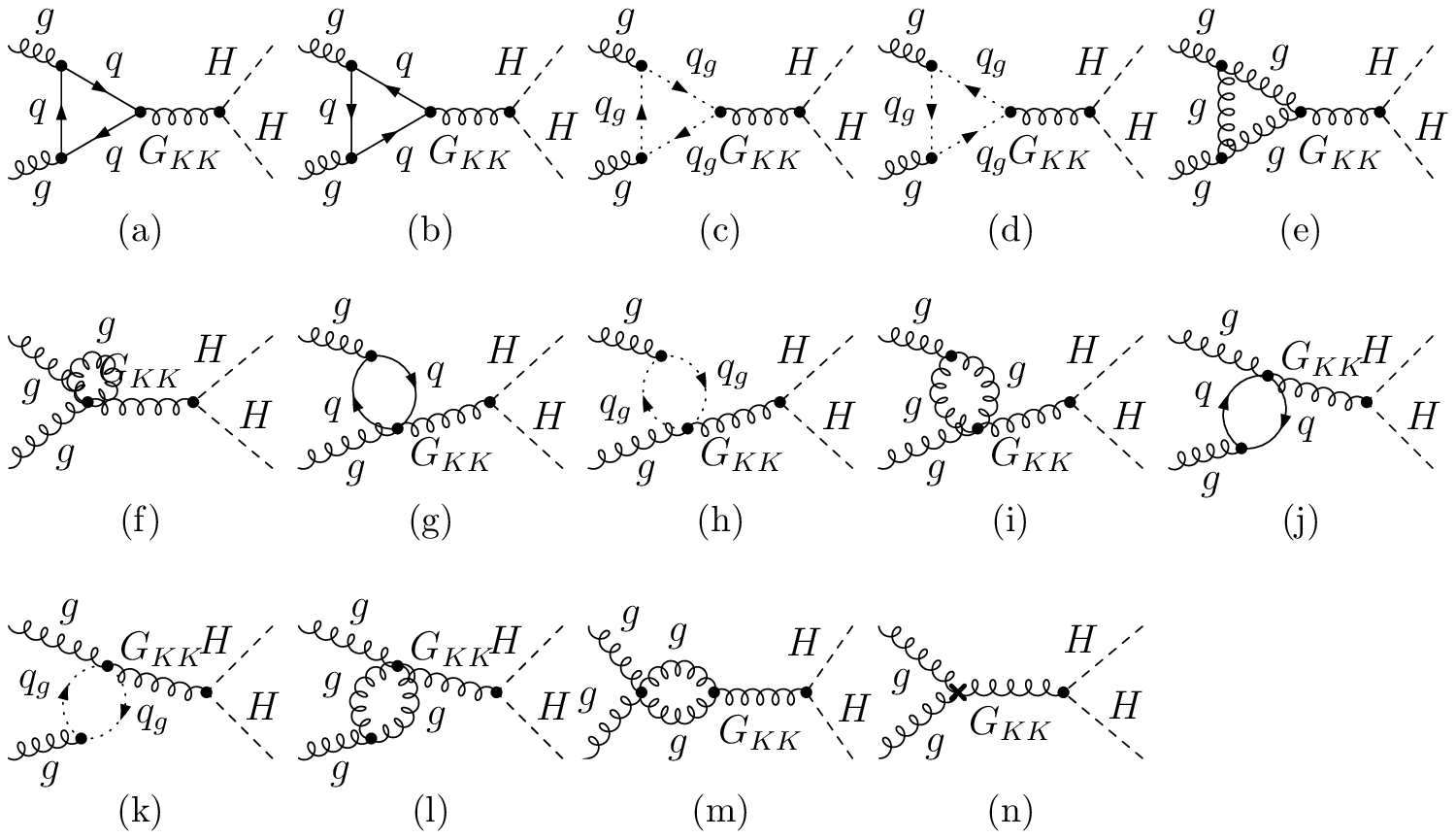}
\vspace{-11.5cm} \caption{\label{fig3} The QCD one-loop Feynman
diagrams for the partonic process $gg\rightarrow G_{KK} \rightarrow
HH$ (a)-(m) and the counter term diagrams(n) correspond to
Fig.\ref{fig1}(e), where q represents u-,d-,c-,s-, b- and t-quark
while $G_{KK}$ represents spin-2 KK graviton.}
\end{figure}
\begin{figure}[hbtp]
\vspace{-5cm} \hspace*{-1.3cm}
\includegraphics[scale=0.8]{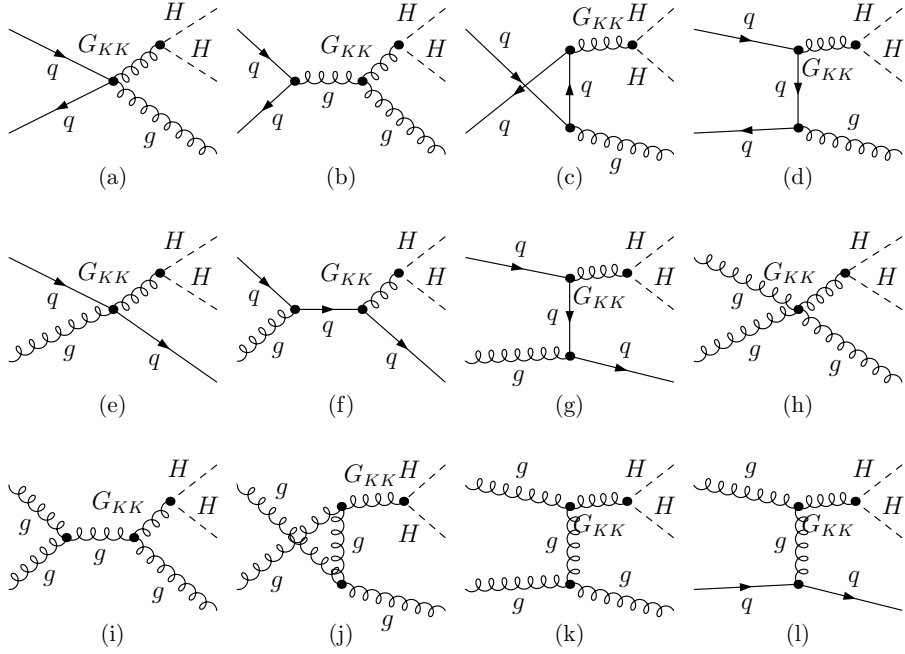}
\vspace{-10.5cm} \caption{\label{fig4} The tree-level Feynman
diagrams for the real gluon/light-(anti)quark emission subprocess
$q\bar{q}\rightarrow G_{KK}\rightarrow HH g$ related to
Eq.\ref{qqemission}(1)[Fig.(a)-(d)], $q(\bar{q})g\rightarrow
G_{KK}\rightarrow HH q(\bar{q})$ related to
Eq.\ref{qqemission}(2,3)[Fig.(e)-(g)], $gg\rightarrow
G_{KK}\rightarrow HH g$  related to
Eq.\ref{ggemission}(1)[Fig.(h)-(k)], $gq(\bar{q})\rightarrow
G_{KK}\rightarrow HH q(\bar(q))$ related to
Eq.\ref{ggemission}(2,3)[Fig.(l)], where q represents u-,d-,c-,s-,
b- and t-quark while $G_{KK}$ represents spin-2 KK graviton. }
\end{figure}

\begin{itemize}
 \item The tree level contributions defined as $\sigma^{LED_{BORN}}_{q\bar{q},gg}$ which
include both the quark-antiquark collision and gluon-gluon fusion
contributions, see Fig.\ref{fig1}(d,e), directly connected by the
effects of KK excitation of gravitons. There are two gravitational
vertices thus such contributions are in the effective order
${\cal{O}}(\kappa^2)$.
 \item The ${\cal{O}}(\alpha_s)$ level one-loop virtual corrections to the leading order process as
well as the renormalization of the leading order cross section which
we define as
$\sigma^{{LED}_{BORN}\otimes{LED}_{LOOP}}_{q\bar{q},gg}$. The QCD
loop level Feynman diagrams of the quark-antiquark collision initial
state part can be found in Fig.\ref{fig2}, totally 4 loop diagrams
and 1 counterterm diagram, while the gluon-gluon fusion
contributions can be found in Fig.\ref{fig3} with 13 loop diagrams
and 1 counterterm diagram as well. These contributions are in the
effective order ${\cal{O}}(\kappa^2\alpha_s)$.
 \item The SM one-loop prediction interference to the leading order process
which defined as $\sigma^{{LED}_{BORN}\otimes{SM}_{LOOP}}_{gg}$.
These contributions are in the order
${\cal{O}}((\sqrt{\kappa}\alpha)\alpha_s)$. We keep these
contributions for a fully consideration and it may also interesting
to see how large are these contributions.
 \item Finally the real gluon and quark emission contributions.
The Feynman diagrams can be found in Fig.\ref{fig4}. The cross
sections are defined as $\sigma_{q\bar{q},gg}^{REAL}$ which are also
in the order ${\cal{O}}(\kappa^2\alpha_s)$. We denote these parts
by:
\begin{eqnarray}\nonumber
&&(1)\ \ q(p_1)\bar{q}(p_2) \rightarrow G_{KK} \rightarrow
H(p_3)H(p_4) g(p_5) \\ \nonumber
&&(2)\ \ q(p_1)g(p_2) \rightarrow G_{KK} \rightarrow H(p_3)H(p_4) q(p_5) \\
&&(3)\ \ g(p_1)\bar{q}(p_2) \rightarrow G_{KK} \rightarrow
H(p_3)H(p_4) (\bar{q}(p_5) \\ \nonumber \label{qqemission}
\end{eqnarray}
and
\begin{eqnarray} \nonumber
&&(1)\ \ g(p_1)g(p_2) \rightarrow G_{KK} \rightarrow H(p_3)H(p_4)
g(p_5) \\ \nonumber
&&(2)\ \ q(p_1)g(p_2) \rightarrow G_{KK} \rightarrow H(p_3)H(p_4) q(p_5) \\
&&(3)\ \ g(p_1)\bar{q}(p_2) \rightarrow G_{KK} \rightarrow
H(p_3)H(p_4) (\bar{q}(p_5) \\ \nonumber \label{ggemission}
\end{eqnarray}
\end{itemize}
refer to emission process related to quark-quark collision and
gluon-gluon fusion sub-processes, respectively.

\par
Finally, we write the total cross section as sum of the above
definitions to the formula below:
\begin{eqnarray}\label{3.3}
 \sigma^{tot}&=& \sigma^{SM} + \sigma^{LED} \\\label{3.4}
             &=&\sigma^{SM_{LOOP}}_{gg}  \\\label{3.5}
             &+&\sigma^{LED_{BORN}}_{q\bar{q}}
              + \underline{ \sigma^{LED_{BORN} \otimes LED_{LOOP}}_{q\bar{q}} + \sigma^{{LED}_{REAL}}_{q\bar{q}}   }_{\Delta\sigma^{LED}_{q\bar{q}}}  \\\label{3.6}
             &+&\sigma^{LED_{BORN}}_{gg}
              + \underline{ \sigma^{LED_{BORN} \otimes LED_{LOOP}}_{gg} + \sigma^{LED_{BORN} \otimes SM_{LOOP} }_{gg}  + \sigma^{{LED}_{REAL}}_{gg}  }
_{\Delta\sigma^{LED}_{gg}}   \\
             &=&\sigma^{SM_{LOOP}}_{gg}+\sigma^{LED}_{q\bar{q}NLO} + \sigma^{LED}_{gg NLO} \\\label{3.8}
             &=&\underline{ \sigma_{gg}^{SM_{LOOP}} + \sigma^{LED}_{q\bar{q} LO} + \sigma^{LED}_{gg LO} }_{\sigma^{LO}} +
 \Delta\sigma^{LED}_{q\bar{q}} + \Delta\sigma^{LED}_{gg}
\end{eqnarray}

\par
The reason to do this is to introduce the symbolic notations we use
later and make it convenient for our numerical calculations.
Eq.\ref{3.3} means we separate the total contributions into two
parts. Eq.\ref{3.4} related to the SM contributions which include
only UV and IR safe terms. Eq.\ref{3.5} and Eq.\ref{3.6} related to
the quark-antiquark collision and gluon-gluon fusion contributions.
Both are summed by two parts. One is the born level contributions
the other are the loop induced LED contributions where we use
$\Delta\sigma^{LED}_{q\bar{q},gg}$ to notate.  We use
$\sigma^{LED}_{q\bar{q} NLO}$ and  $\sigma^{LED}_{gg NLO}$ to keep
notation of Eq.\ref{3.5} and \ref{3.6}, respectively. Finally, we
define the SM prediction plus the leading order LED contributions as
$\sigma^{LO}$ and the other terms are the shifted terms coming from
the loop induced LED effects, see Eq.\ref{3.8}.

\par
There exist UV and soft/collinear IR singularities in the
calculations of $\sigma^{LED_{BORN} \otimes
LED_{LOOP}}_{q\bar{q},gg}$. To remove the UV divergences, we need
only the wave function renormalization constants for the quark and
gluon fields. We introduce the renormalization constants $\delta
Z_{\psi_{q,L,R}}$ for massless quark (q=u,d,c,s,b) fields and
$\delta Z_{A}$  for the gluon field defined as
\begin{eqnarray}
\psi^{0}_{q,L,R}=(1+\delta Z_{\phi_{q,L,R}})^{\frac{1}{2}}
\psi_{q,L,R}, \ \ \  A^{a0}_{\mu}=(1+\delta Z_A)^{\frac{1}{2}}
A^a_{\mu}
\end{eqnarray}

\par
In the modified minimal subtraction ($\overline{MS}$)
renormalization scheme the renormalization constants for the
massless quarks are expressed as
\begin{eqnarray}
\delta Z_{\psi_{q,L}} &=&
-\frac{\alpha_{s}}{4\pi}C_{F}(\Delta_{UV}-\Delta_{IR}),
~~~\delta Z_{\psi_{q,R}} = -\frac{\alpha_{s}}{4\pi}C_{F}\left(\Delta_{UV}-\Delta_{IR}\right),  \\
\delta Z_{A} &=&
\frac{\alpha_{s}}{4\pi}\left(\frac{5}{3}C_{A}-\frac{4}{3}n^{UV}_{f}T_{F}\right)\Delta_{UV}
+\frac{\alpha_{s}}{4\pi}\left(\frac{5}{3}C_{A}-\frac{4}{3}n^{IR}_{f}T_{F}\right)\Delta_{IR},
\end{eqnarray}

\par
In the above equations $\mu_r$ is the renormalization scale,
$C_{F}=\frac{4}{3}$, $C_{A}=3$, $T_{F}=\frac{1}{2}$, $n^{UV}_{f}=6$
corresponds to the six flavor quarks ($u$, $d$, $c$, $s$, $t$, $b$),
whereas $n^{IR}_{f}=5$ is the number of the massless quarks ($u$,
$d$, $s$, $c$, $b$). Moreover, $\Delta_{UV}=
\frac{1}{\epsilon_{UV}}\Gamma
(1+\epsilon_{UV})(4\pi)^{\epsilon_{UV}}$ and
$\Delta_{IR}=\frac{1}{\epsilon_{IR}}\Gamma
(1+\epsilon_{IR})(4\pi)^{\epsilon_{IR}}$ refer to the UV and IR
divergences, respectively.

\par
By adding renormalization part to the virtual corrections, any
ultraviolet(UV) singularities are regulated. Divergences arising
from soft gluon emission removed by combining virtual and real
emission corrections $\sigma^{LED_{REAL}}_{q\bar{q},gg}$.
Singularities associated with initial state collinear gluon emission
are absorbed into the definition of the parton distribution
functions. We employ the $\overline{MS}$ scheme for the parton
distributions functions. Similar to the virtual part, we utilize
dimensional regularization to control the singularities of the
radiative corrections, which are organized using the two cutoff
phase space slicing(TCPSS) method\cite{2PSS:Owens}. We adopt TCPSS
to isolate the IR singularities by introducing two cutoff parameters
$\delta_{s}$ and $\delta_{c}$. An arbitrary small $\delta_{s}$
separates the three-body final state phase space into two regions:
the soft region ($E_{5}\leq \delta_{s}\sqrt{\hat{s}}/2$) and the
hard region ($E_{5}>\delta_{s}\sqrt{\hat{s}}/2$). The $\delta_{c}$
separates hard region into the hard collinear ($HC$) region and hard
noncollinear ($\overline{HC}$) region. The criterion for separating
the HC region is described as follows: the region for real
gluon/light-(anti)quark emission with $\hat{s}_{15}$ (or
$\hat{s}_{25}$) $< \delta_{c}\hat{s}$ (where
$\hat{s}_{ij}=(p_{i}+p_{j})^{2}$) is called the $HC$ region.
Otherwise it is called the $\overline{HC}$ region. Then the cross
section for each of the real emission partonic processes can be
written as
\begin{equation}
 \hat{\sigma}^R=\hat{\sigma}^{S}+\hat{\sigma}^{H}=\hat{\sigma}^{S}+\hat{\sigma}^{HC}+\hat{\sigma}^{\overline{HC}}.
\end{equation}

\par
After combining all these contributions above, the UV and IR
singularities in $\sigma^{tot}$ are exactly canceled. Dependence on
the arbitrary small cutoff parameters $\delta_{s}$ and $\delta_{c}$
are exactly vanished. These cancelations can be verified numerically
in our numerical calculations. We use Feynarts to create the
amplitudes and FormCalc to perform the numerical calculation. To
compute the one-loop tensor integrals we use our modified LoopTools
\cite{FeynArts,FormCalc,LoopTools} which supply an automatic
numerical checks for the Inferared(IR) divergence and cross check
with OneLoop\cite{Oneloop} at enough random phase space points.

\section{Numerical results and discussions}
\par
In the SM frame, given the small total cross section, it is clear
that even for $\sqrt{s}=14$ TeV and a target luminosity of $\cal
O$($1000 fb^{-1}$) one need to focus on the Higgs decay channels
with the largest branching ratios to visible final states to observe
$pp\rightarrow HH+X$ such as $H\rightarrow b\bar{b}(59.48\%)$,
$H\rightarrow W^-W^+(20.78\%)$ , $H\rightarrow
\tau^-\tau^+(\mu^-\mu^+)(6.12\%)$\cite{HdecayBranch}. A feasibility
study on the decay channels $HH\rightarrow b\bar{b}\gamma\gamma$,
$HH\rightarrow b\bar{b}\mu^{-}\mu^{+}$, finding that with 600
$fb^{-1}$ one expects 6 signal and 11 background events, giving s
significance of about $1.5\sigma$\cite{HH2bbrrbbmm}. For the channel
$HH\rightarrow b\bar{b}W^-W^+\rightarrow b\bar{b}l \nu jj$,
Ref\cite{HH2bbtt} claimed with 57 signal and 119 background events
at $600 fb^{-1}$ by employing new techniques and assuming good
$\tau$ reconstruction efficiency($\sim 80\%$), thus make such
channel a promising one. As can be seen, though challenge, Higgs
pair production in SM can be measured. What's more, we hope to
concentrate on the Higgs pair production under LED frame and find
out whether this production can be enhanced and given more chance to
be tested.

\par
In the numerical calculations, we take the input parameters as
$M_Z=91.1876~{\rm GeV}$, $M_W=80.399~{\rm
GeV}$\cite{ParticleDataGroup}, $\alpha(M_Z^2)^{-1}=1/127.934$,
$M_{H}=125~{\rm GeV}$\cite{SMHiggs125GeV_ATLAS, SMHiggs125GeV_CMS}.
The factorization and renormalization scales are chosen to be the
same $\mu_r=\mu_f=\mu_0=M_{H}=125~{\rm GeV}$. Numerical analysis is
done at the colliding c.m.s. energy 8TeV and 14 TeV LHC for the
early and future LHC.

The UV and IR safeties are verified numerically after combining all
the contributions at the QCD loop level. To check the UV and IR
divergence cancelation, we display a random phase space point as
well as the cancelation for different divergent parameters, see in
Table\ref{tab1}. One thing that should be emphasized is
$\sigma^{LED_{BORN} \otimes LED_{LOOP}
}_{q\bar{q},gg}+\sigma^{LED_{BORN} \otimes SM_{LOOP} }_{gg}$ should
include the conterterm contributions as well as the soft and
collinear singularity terms coming from the hard emission. We
implement this into our monte carlo codes which supply an automatic
check of the dependence on these divergence parameters. We can see
the UV and IR divergence can be canceled at high precision level in
all the phase space, thus leading the continuance of our following
calculation.

\begin{table}[hbtp]
\begin{center}
\begin{tabular}{|c|}
\hline
A Random Phase Space Point for \\
$ pp\rightarrow i(p_1) j(p_2) \rightarrow H(p_3) H(p_4),\ \ ij=(q\bar{q},gg) $ \\
\hline
$p_1$=(152.42677337684606         0        0        152.42677337684606)\\
$p_2$=(152.42677337684606         0        0       -152.42677337684606)\\
$p_3$=(152.42677337684606        86.461267952196764    0        11.548609707441768)\\
$p_4$=(152.42677337684606       -86.461267952196764    0       -11.548609707441768)\\
\hline
\end{tabular}
\begin{tabular}{| l | c | c |}
    \hline
divergence & \small $\sigma^{LED_{BORN}}_{q\bar{q},gg}$ &\small
$\sigma^{LED_{BORN} \otimes LED_{LOOP}
}_{q\bar{q},gg}+\sigma^{LED_{BORN} \otimes SM_{LOOP} }_{gg}$[fb] \\
\hline
\small $\frac{1}{\epsilon_{UV}}=\frac{1}{\epsilon^2_{IR}}=\frac{1}{\epsilon_{IR}}=0$       & 0.8461630560055 & 0.1259656696\ 911 \\
\small $\frac{1}{\epsilon_{UV}}=\frac{1}{\epsilon^2_{IR}}=\frac{1}{\epsilon_{IR}}=10^{10}$ & 0.8461630560055 & 0.1259656696\ 636 \\
    \hline
\end{tabular}
\end{center}
\vspace*{-0.8cm}
\begin{center}
\begin{minipage}{14cm}
\caption{\label{tab1} The UV and IR divergence cancelation at one
given random phase space point for the loop contribution. Notice
that in the loop terms, we include the conterterm contribution as
well as the soft and collinear singularity terms coming from the
hard emission contributions. }
\end{minipage}
\end{center}
\end{table}

\par
Since the total cross section is independent of the soft cutoff
$\delta_s(=\Delta E_g/E_b, E_b=\sqrt{\hat{s}}/2)$ and the collinear
cutoff $\delta_c$, we display the curves of the loop induced QCD
corrections to integrated cross section for $pp\rightarrow ij
\rightarrow G_{KK} \rightarrow HH+X$ process versus the cutoff
$\delta_s$ at the $\sqrt{s}$ = 14 TeV LHC in the LED model, where we
take $\mu_f=\mu_r=\mu_0=M_H$, $M_s$ = 3.5 TeV , $\delta$ = 3 and
$\delta_c=\delta_s/100$. For the strong coupling constant
$\alpha_s(\mu)$, we use the two-loop evolution of it with the QCD
parameter $\Lambda^{n_f=5}$ = 226 MeV and get $\alpha_s(\mu_0)$ =
0.113. $N_f$ is the number of the active flavors. Some of the LED
results are listed in Table 2. The terms gg, uu, dd refer to pure
gluon-gluon fusion, $u\bar{u}+c\bar{c}$,
$d\bar{d}+s\bar{s}+b\bar{b}$ collisions respectively. We show the
$\delta_s$ dependence on all the sub-contribution terms. It is shown
clearly that the NLO QCD correction
($\Delta\sigma^{LED}_{q\bar{q},gg NLO}$) does not depend on the
arbitrarily chosen values of $\delta_s$ and $\delta_c$ within the
calculation errors. In the further numerical calculations, we fix
$\delta_s = 10^{-4}$ and $\delta_c=\delta_s/100$. To satisfy the
unitary constraint, we adopt the cut $\sqrt{\hat{s}} < M_s$ for the
whole phase space.

\begin{table}
\begin{center}
\begin{tabular}{l c c c c c r}
\hline\hline
  $\delta_s$    &&\multicolumn{5}{c}{$\Delta\sigma^{LED}_{NLO}$[pb]} \\ [0.5ex]
                  && $uu$ && $dd$  && $gg$ \\
\hline
$5\times 10^{-3}$ &&7.843304975113E-03  &&3.524461919262E-03   &&1.946637625053E-02    \\
$1\times 10^{-3}$ &&7.846245116613E-03  &&3.522936306130E-03   &&1.943421303121E-02    \\
$5\times 10^{-4}$ &&7.846382881667E-03  &&3.528078248755E-03   &&1.945856138503E-02    \\
$1\times 10^{-4}$ &&7.846562713918E-03  &&3.534350158736E-03   &&1.942754069450E-02    \\
$5\times 10^{-5}$ &&7.844722993687E-03  &&3.532865728080E-03   &&1.936325027843E-02    \\
$1\times 10^{-5}$ &&7.793411078227E-03  &&3.534430130933E-03   &&1.933905866994E-02    \\
\hline\hline
\end{tabular}
\end{center}
\vspace*{-0.8cm}
\begin{center}
\begin{minipage}{14cm}
\caption{\label{tab2} The dependence of the loop induced QCD
correction to the integrated cross section for the $pp\rightarrow ij
\rightarrow G_{KK} \rightarrow HH+X(ij=q\bar{q},gg)$ at the
$\sqrt{s}$ = 14TeV LHC in the LED model, where we set
$\mu_r=\mu_f=\mu_0=M_H$, $M_s=3.5TeV$, $\delta=3$ and
$\delta_c=\delta_s/100$. The terms gg, uu, dd refer to pure
gluon-gluon fusion, $u\bar{u}+c\bar{c}$,
$d\bar{d}+s\bar{s}+b\bar{b}$ collisions respectively.}
\end{minipage}
\end{center}
\end{table}

\begin{figure}[hbtp]
\centering
\includegraphics[scale=0.7]{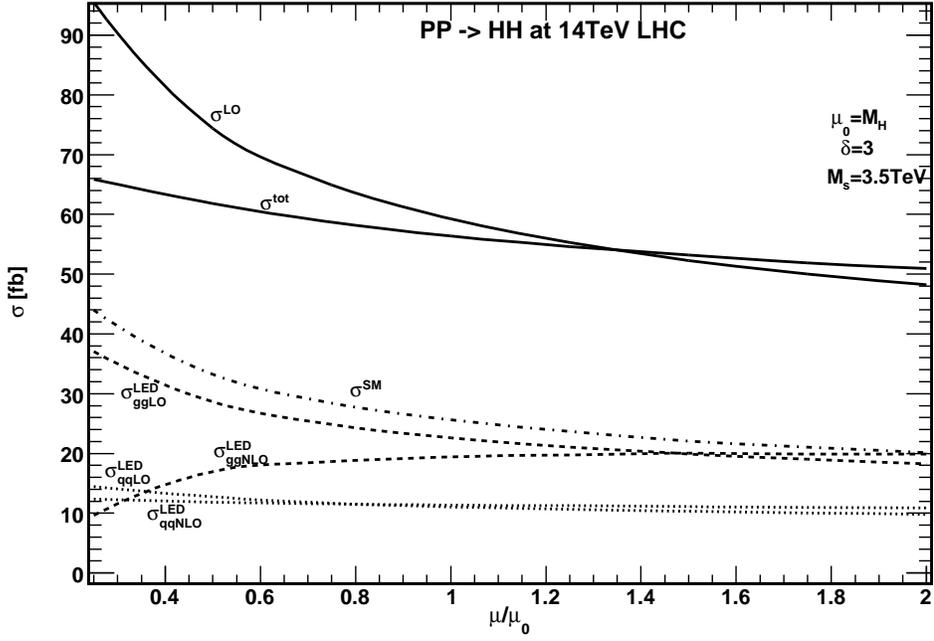}
\caption{\label{fig5}  The scale($\mu$) dependence of the loop
induced SM contribution ($\sigma^{SM}$)[dash-dotted line], leading
order born ($\sigma^{LO}$) and QCD loop corrected total
($\sigma^{tot}$)[solid line] cross sections as well as their
seperate sub-contributions ($\sigma^{LED}_{q\bar{q},gg}$)[dotted and
dashed lines] in the SM and LED model at the $\sqrt{s} = 14$ TeV LHC
with $\mu=M_H$, $M_s = 3.5$ TeV and $\delta$ = 3.
 }
\end{figure}

\par
In Fig.\ref{fig5}, we show the scale($\mu$) dependence of the loop
induced SM contribution ($\sigma^{SM}$), leading order born
($\sigma^{LO}$) and QCD loop corrected total ($\sigma^{tot}$) cross
sections as well as their seperate sub-contributions
($\sigma^{LED}_{q\bar{q},gg}$) in the SM and LED model at the
$\sqrt{s} = 14$ TeV LHC, and we define the corresponding K-factor as
$k_{\mu}=\sigma^{tot}/\sigma^{LO}$. There we take the input as $M_s
= 3.5$ TeV and $\delta$ = 3. From this figure, we can see:
\begin{itemize}
 \item For the SM prediction $\sigma^{SM}$, see the dot-dashed line in Fig.\ref{fig5}, though coming from the
SM loop diagrams(Fig.\ref{fig1}(a,b,c)), its cross section changes
from 44 fb to 20 fb if the scale $\mu$ goes from 0.2 to 2 $\mu_0$,
which implies obvious $\mu$ dependence.
 \item For the pure LO and QCD loop induced LED effects, contributions from the gluon fusion ($\sigma^{LED}_{gg NLO}$) is about
two times of it coming from quark-antiquark collision
($\sigma^{LED}_{q\bar{q} NLO}$). Separately, the $\sigma^{LED}_{ij
NLO}(ij=q\bar{q},gg)$ reduce the $\mu$ dependence on the
corresponding $\sigma^{LED}_{ij LO}$ contributions, see the dotted
and dashed lines in Fig.\ref{fig5} for more details.
 \item Sum all the contributions together, we get the $\mu$ dependence on the total cross section
$\sigma^{LO}$ and $\sigma^{tot}$ with the definition in
Eq.\ref{3.3}, see the solid lines in Fig.\ref{fig5}. We can see the
cross section value in the LED model varies from  95.4(65.9) fb to
48.2(50.9) fb for the LO and NLO contribution when $\mu$ goes from
0.2 $\mu_0$ to 2 $\mu_0$ at the $\sqrt{s}$ = 14 TeV LHC. So that the
K-factor in the LED model varies from 1.44871 to 0.946948. We see
from this panel that the NLO QCD corrections in the LED model
totally reduce the factorization/renormalization scale uncertainty
obviously. In further calculations we fix $\mu=\mu_0=M_H$.
\end{itemize}

\begin{figure}[hbtp]
\centering
\includegraphics[scale=0.6]{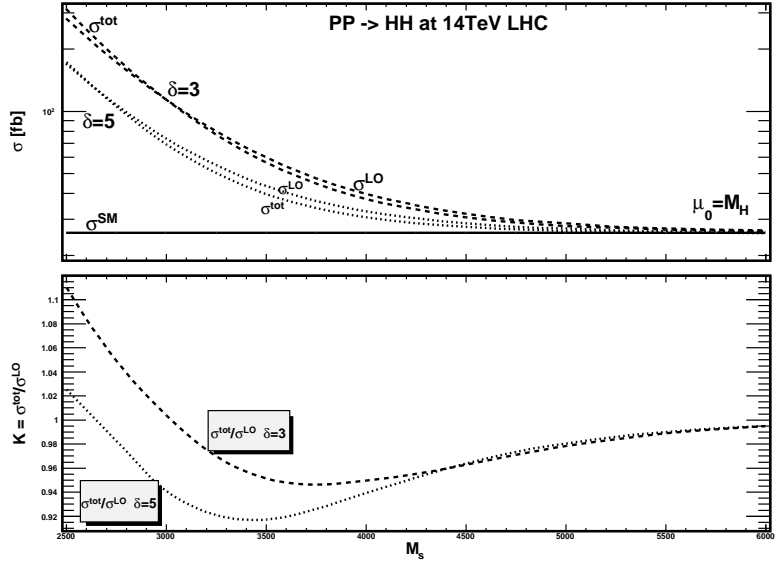}
\caption{\label{fig6} The cross sections and the
K-factor($\sigma^{tot}/\sigma^{LO}$) for the process $pp\rightarrow
ij \rightarrow G_{KK} \rightarrow HH+X(ij=q\bar{q},gg)$ in the
SM($\sigma^{SM}$) and LED model at both the born($\sigma^{LO}$)
level and loop induced level($\sigma^{tot}$) as functions of $M_s$
with $\mu_0=M_H=125$ GeV at the $\sqrt{s}=14$ TeV future LHC. The
solid line refer to the SM prediction while the dotted and dashed
line correspond to $\delta=3$ and $\delta=5$, respectively. }
\end{figure}

\par
\begin{figure}[hbtp]
\centering
\includegraphics[scale=0.6]{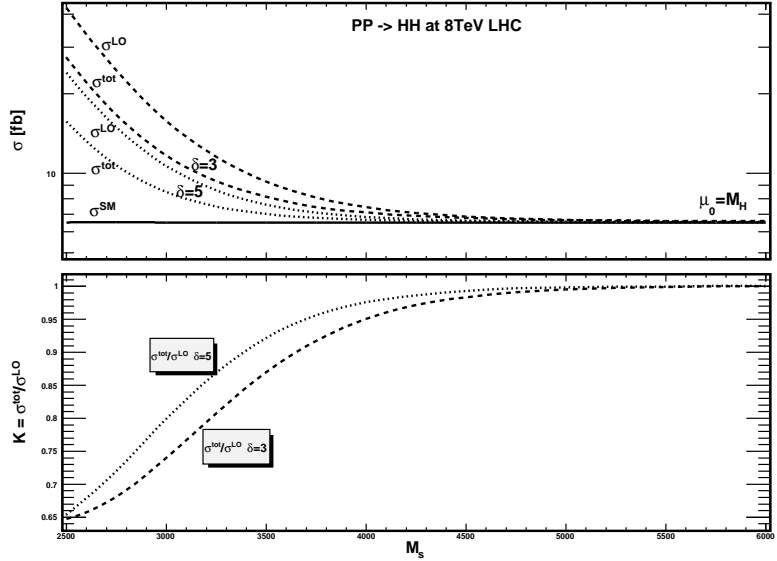}
\caption{\label{fig7} The cross sections and the
K-factor($\sigma^{tot}/\sigma^{LO}$) for the process $pp\rightarrow
ij \rightarrow G_{KK} \rightarrow HH+X(ij=q\bar{q},gg)$ in the
SM($\sigma^{SM}$) and LED model at both the born($\sigma^{LO}$)
level and loop induced level($\sigma^{tot}$) as functions of $M_s$
with $\mu_0=M_H=125$ GeV at the $\sqrt{s}=8$ TeV early LHC. The
solid line refer to the SM prediction while the dotted and dashed
line correspond to $\delta=3$ and $\delta=5$, respectively. }
\end{figure}

In upper panel of Fig.\ref{fig6}, we depict the cross sections for
the process $pp\rightarrow ij \rightarrow G_{KK} \rightarrow
HH+X(ij=q\bar{q},gg)$ in the SM($\sigma^{SM}$) and LED model at both
the tree($\sigma^{LO}$) level and NLO QCD loop induced
level($\sigma^{tot}$) as functions of the fundamental scale $M_s$
from 2.5 TeV to 6 TeV with  $\mu_0=M_H=125$ GeV at the $\sqrt{s}=14$
TeV future LHC. The solid line presents the SM
prediction($\sigma^{SM}$) while the dotted and dashed line
correspond to the total LED effects at both the born($\sigma^{LO}$)
level and loop induced level($\sigma^{tot}$) with the extra
dimension number $\delta$ being 3 and 5, respectively. Notice
$\sigma^{tot}=\sigma^{SM}+\sigma^{LED}$ and $\sigma^{LO}$ are
defined in Eq.\ref{3.3} and Eq.\ref{3.8} respectively. From the
figures one finds that the largest deviation from the SM due to LED
occurs at small values of $M_s$ and $\delta$. With the fixing value
of $\delta$, the born level LED effects and the QCD loop induced LED
effects deviate at different value of $M_s$ behavior as the dotted
and dashed lines presented. In order to compare the relative size of
such deviation, we depict the K-factor($\sigma^{tot}/\sigma^{LO}$)
in the lower panel in Fig.\ref{fig6}. One can see for $\delta=3$,
the K-factor is larger than 1 when $M_s<3$ TeV with the highest
value 1.13, while $M_s>3$ TeV the K-factor can turn to it's lowest
value 0.95 which means the loop effects reduce the born effects
about 5$\%$. For $\delta=5$, the K-factor is always less than 1 if
$M_s>2.6$ TeV. When $M_s>4.5$TeV, both line close to the the small
value of 0.98. The same thing has been depicted in Fig.\ref{fig7}
for the 8 TeV early LHC. We take the same input parameters as above,
we found that though the LED effects enhance the SM prediction
generally, the total loop induced effects reduce the born level LED
effects. Their deviation become large if $M_s$ become small and the
K-factor turn to their lowest value 0.65 with $M_s=2.5$ TeV. When
$M_s$ become larger then 5 TeV, no matter the value of $\delta$, the
K-factor turns to 1 which means the born level and loop induced LED
effects become close to each other and also close to their SM
predictions.

\begin{figure}[hbtp]
\vspace{-1.2cm} \centering
\includegraphics[scale=0.36]{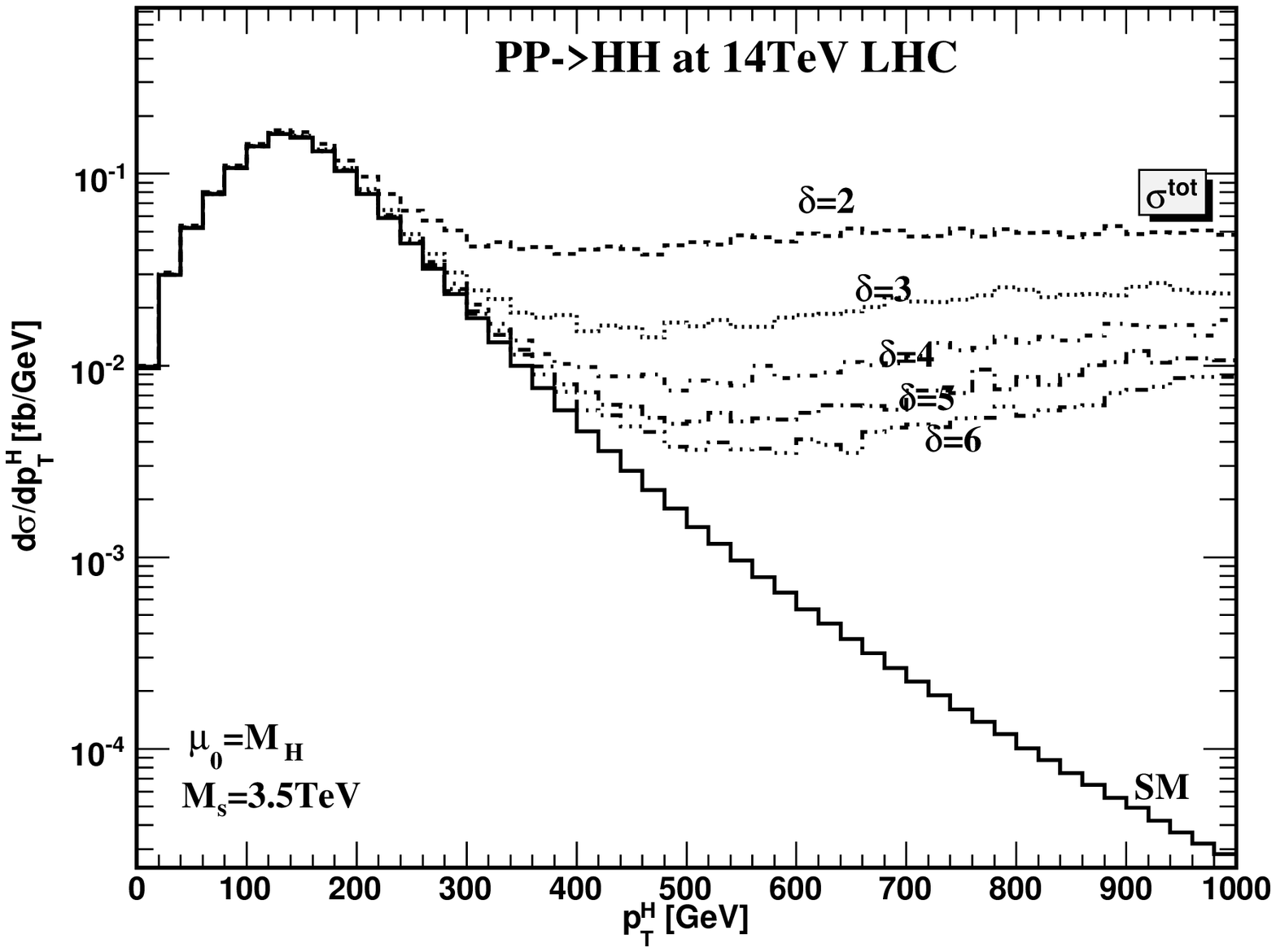}
\includegraphics[scale=0.36]{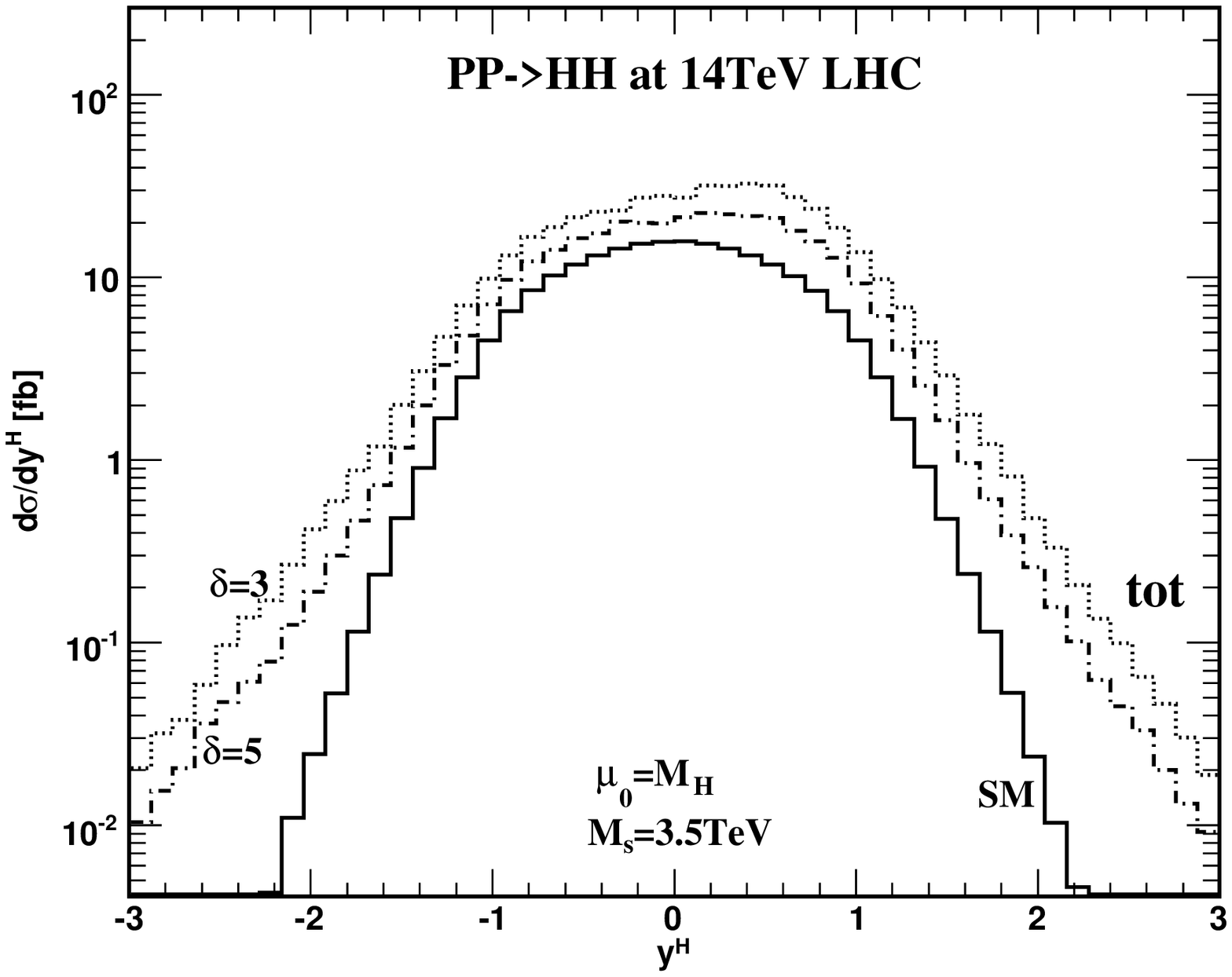}
\caption{\label{fig8}  The transverse momentum($p_T$)[left panel]
and Rapidity($y$)[right panel] distribution of Higgs bosons for the
process $pp\rightarrow ij \rightarrow G_{KK} \rightarrow
HH+X(ij=q\bar{q},gg)$ at 14 TeV LHC with $M_s=3.5TeV$ and
$\mu=\mu_0=M_H$. The solid line presents the SM prediction while the
dashed, dotted, dash-dotted and dash-dash-dotted and
dash-dash-dot-dotted lines refer to $\delta=2,3,4,5,6$,
respectively. }
\end{figure}

\par
In Fig.\ref{fig8}, we show the transverse momentum($p_T$)[left
panel] and the rapidity($y$)[right panel] distributions of the Higgs
boson for the process $pp\rightarrow ij \rightarrow G_{KK}
\rightarrow HH+X(ij=q\bar{q},gg)$ in the SM($d\sigma^{SM}/dp_T^{H}$,
$d\sigma^{SM}/dy^{H}$) and LED model($d\sigma^{tot}/dp_T^{H}$,
$d\sigma^{tot}/dy^{H}$) at the 14 TeV LHC. There the results are for
$M_s=3.5$ TeV at the fixed value 3 for the number of extra
dimensions and obtained by taking the input parameters mentioned
above. The LED effects gently raise the SM prediction at values of
high $p^H_T$ regions. Rapidity distribution is defined as
$\frac{d\sigma}{d\eta}$ with $\eta=\frac{1}{2} \textbf{ln}(P_1\cdot
q)/(P_2\cdot q)$, where $P_1$ and $P_2$ are incoming proton momenta
and q is the related Higgs boson 4-momenta. As we can see, the
rapidity distributions in the LED model show significantly narrow
peaks around $y = 0$, which implies the large contributions at high
$p_T$ region.

\begin{figure}[hbtp]
\centering
\includegraphics[scale=0.35]{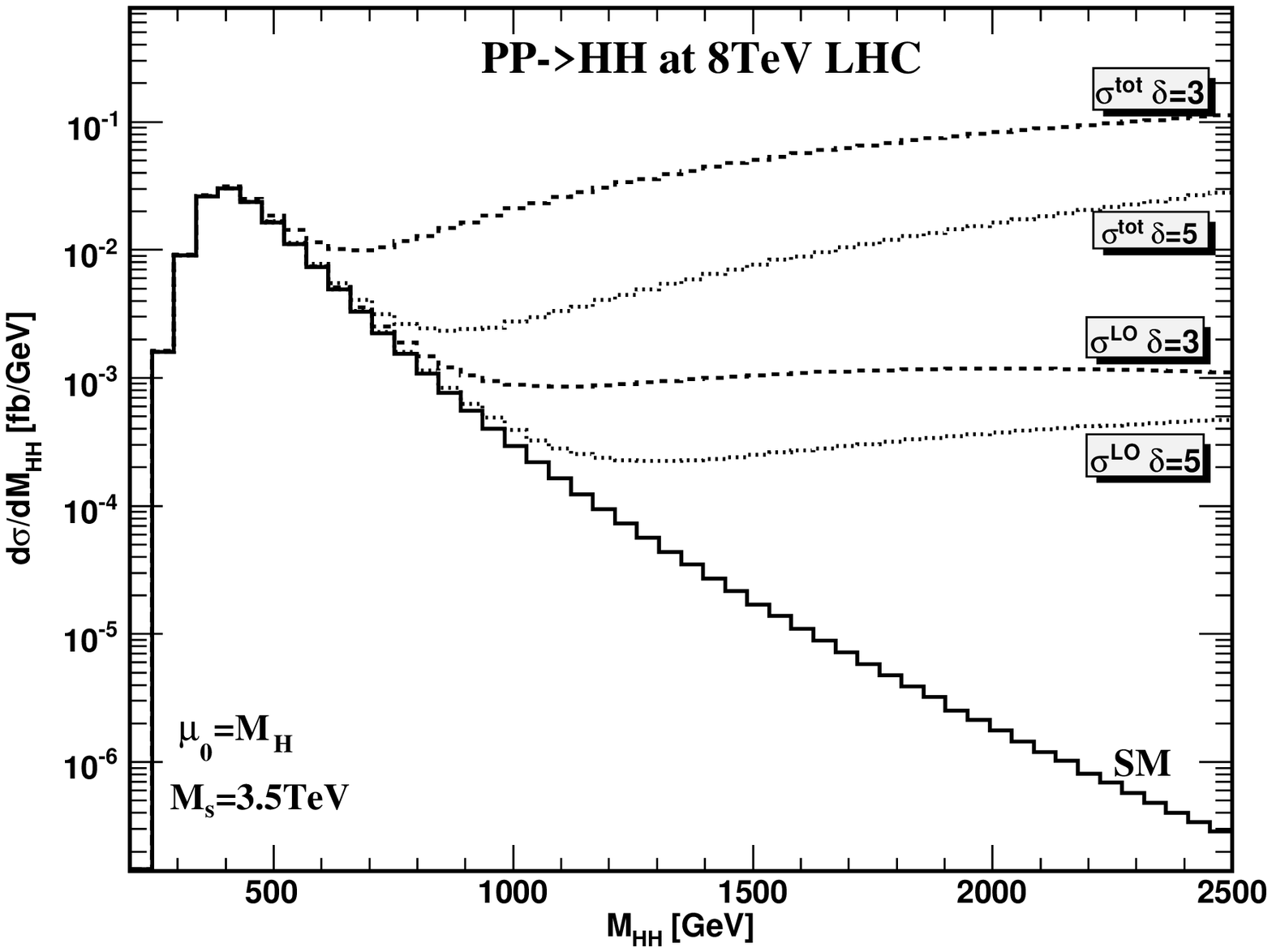}
\includegraphics[scale=0.35]{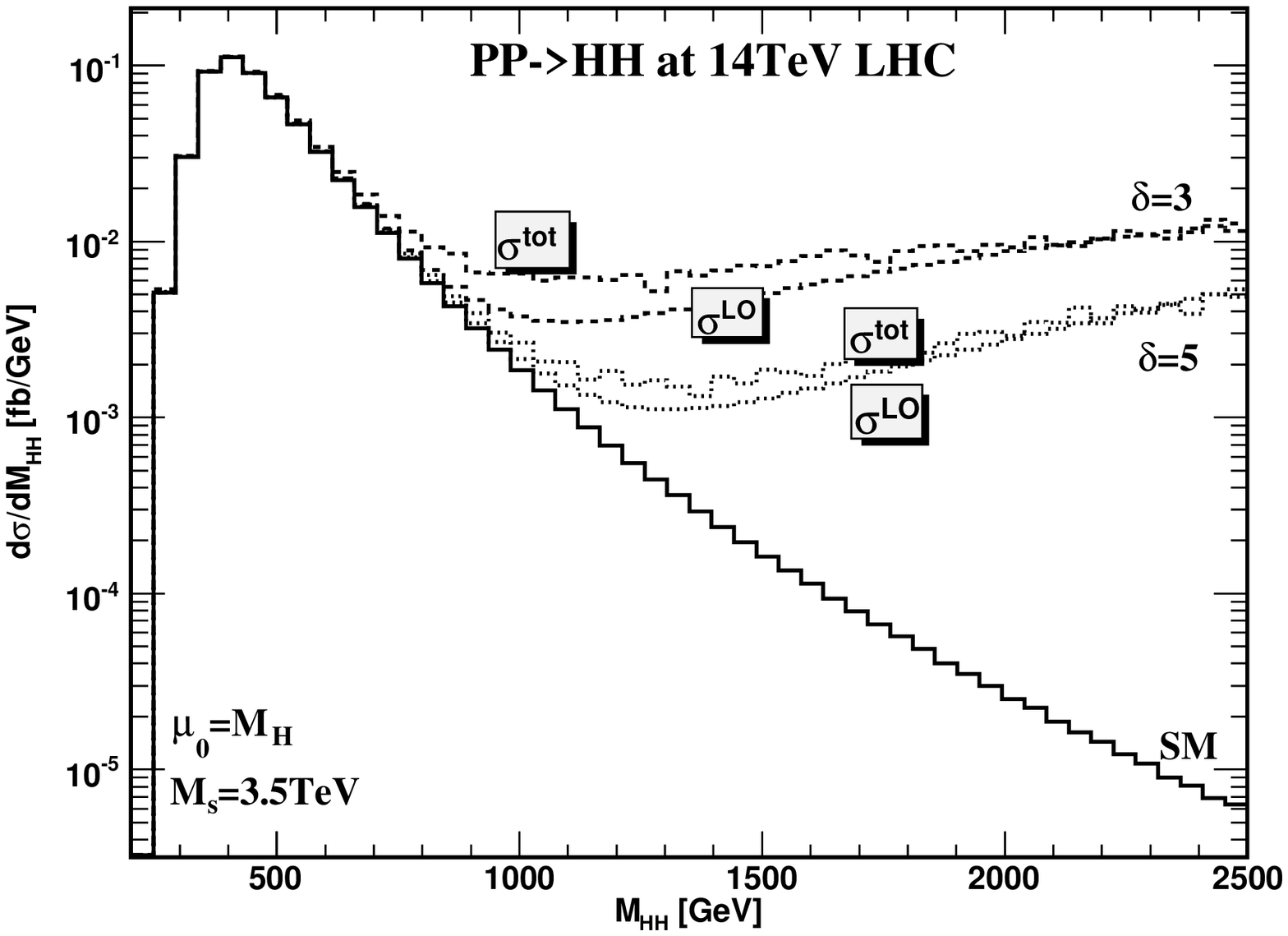}
\caption{\label{fig9}  The invariance mass($M_{HH}$) distribution of
Higgs boson pair for the process $pp\rightarrow ij \rightarrow
G_{KK} \rightarrow HH+X(ij=q\bar{q},gg)$ at 8TeV[left panel] and 14
TeV[right panel] early and future LHC with $M_s=3.5TeV$ and
$\mu=M_H$. The solid line presents the SM prediction while the
dashed, dotted lines refer to $\delta=3,5$, respectively. The upper
one refer to $\sigma^{tot}$ while the lower one refer to
$\sigma^{LO}$. }
\end{figure}

\par
The distributions of the invariance mass($M_{HH}$) distribution of
Higgs boson pair for $pp\rightarrow ij \rightarrow G_{KK}
\rightarrow HH+X(ij=q\bar{q},gg)$ has been displayed in
Fig.\ref{fig9} at the 8 TeV[left panel] and 14 TeV[right panel]
early and future LHC. Same conclusion can be obtained that the LED
effects enhance the distribution at the high $M_{HH}$ region since
in these regions the LED effect dominant the total (SM+LED)
distribution, as a result, more KK modes contribute with the
increase of $M_{HH}$. The lower the dimension is, the higher value
of $M_{HH}$ deviation from its SM prediction as presented in
Fig.\ref{fig9}. It will be also very interesting to compare the
results at 8 TeV with that at 14 TeV LHC. We can see that though 14
TeV LHC has much higher c.m.s. energy than 8 TeV LHC, the
theoretical amplitude at higher c.m.s. LHC is hampered by the
unitary constraint $\sqrt{\hat{s}}<{\rm M_s}$, thus the LED effect
is much suppressed at 14TeV LHC than 8 TeV LHC. Besides, at the 8
TeV LHC, the total contribution $\sigma^{tot}$ deviate from its
leading order contribution $\sigma^{LO}$ obviously. The deviations
become much larger compare to the 14 TeV results. While at 14 TeV
LHC, the deviations only appear in the middle of $M_{HH}$ region
between $\sigma^{tot}$ and $\sigma^{LO}$ when the dimension is fixed
to be the same value.

As have been told, the distribution shows the crutial character that
the LED effects are mainly collected at high $p_T^H$ and $M_{HH}$
region while the SM backgrounds are leaving in the lower regions. We
thus apply two strict cuts $p_T(H)\geq 500$ GeV and $M_{HH}\geq 1$
TeV to strongly suppress the SM backgrounds while leaving the LED
effects almost unchanged. We concentrate on two channels:
$pp\rightarrow HH\rightarrow b\bar{b}\gamma\gamma$ and
$pp\rightarrow HH\rightarrow b\bar{b}\mu^-\mu^+$. The main
backgrounds come from $pp\rightarrow ZZ(ZH)$( with $Z\rightarrow
b\bar{b}(\mu^-\mu^+)$ can be a background to $h\rightarrow
b\bar{b}(\mu^-\mu^+)$) and $pp\rightarrow t\bar{t}\rightarrow
b\bar{b}\mu^-\mu^+\nu\bar{\nu}$. By taking into account all these
backgrounds and adopting the kinematic cuts mensioned above, we
display the signal background ratio $\frac{S}{\sqrt{B}}$ in
Table\ref{tab3} and Table\ref{tab4} corresponding to the two
different decay channels with $M_S$ and $\delta$ as running
parameter and luminosity equal (100, 600)[$fb^{-1}$] at 14 TeV
future LHC. From these two tables we can see that Higgs pair
production in LED model can give significant signal to background
ratio when the scale $M_s\leq5$ TeV and dimension $\delta\leq5$. By
using the $b\bar{b}\gamma\gamma$ channel, one can test LED effects
at $M_s=5$ TeV and $\delta=4$ with $\frac{S}{\sqrt{B}}>3$ while for
$b\bar{b}\mu^-\mu^+$ channel with $\frac{S}{\sqrt{B}}\geq 5$. Both
channels shows good character in testing LED effects. In the tables,
$\sim(<)1$ means the $\frac{S}{\sqrt{B}}$ is close to(less than) 1,
however, this can be improved if we apply the cuts more strict or
wait for the luminosity to be higher at the LHC. It will be
interesting that $pp\rightarrow G_{KK}\rightarrow HH$ can also be a
promising process in testing LED effects.

\begin{table}
\begin{center}
\begin{tabular}{l c c c c c c c c c c c c c r}
\hline\hline
\centering$\frac{S}{\sqrt{B}}$       & \multicolumn{14}{c}{ $b\bar{b}\gamma\gamma$ channel \textcircled{a} 14 TeV with $\cal L$=(100, 600)[$fb^{-1}$]} \\ [0.5ex]
$M_s$[TeV]&\multicolumn{2}{c}{$\delta=2$}&&\multicolumn{2}{c}{$\delta=3$}&&\multicolumn{2}{c}{$\delta=4$}&&\multicolumn{2}{c}{$\delta=5$}&&\multicolumn{2}{c}{$\delta=6$}\\
\hline
3.5   & 26    & 64      && 12.4 & 30   && 11   &26     && 8.2  &20   && 6.5  & 16  \\
4.0   & 11.2  & 27.4    && 6.4  &15.6  && 4.4  &11     && 3.1  &7.5  && 2.6  & 6.3  \\
4.5   & 5.2   & 13      && 2.8  & 6.8  && 1.9  &4.5    && 1.4  &3.3  && 1.1  & 2.6  \\
5.0   & 2.5   & 6.1     && 1.8  &4.2   && 1.3  &3.2    && $\sim 1$  &1.5  && $\sim 1$  &1.2  \\
5.5   & 1.3   & 3.2     && $\sim 1$ & 1.5  && $<1$  &1      && $<1$  &$\sim 1$  && $<1$  &$\sim 1$  \\
\hline\hline
\end{tabular}
\end{center}
\vspace*{-0.8cm}
\begin{center}
\begin{minipage}{14cm}
\caption{\label{tab3} $\frac{S}{\sqrt{B}}$ for the cahnnel $pp\rightarrow HH\rightarrow b\bar{b}\gamma\gamma$
after taking into account all backgrounds and adopt the kinematic cuts. }
\end{minipage}
\end{center}
\end{table}

\begin{table}
\begin{center}
\begin{tabular}{l c c c c c c c c c c c c c r}
\hline\hline
\centering$\frac{S}{\sqrt{B}}$       & \multicolumn{14}{c}{ $b\bar{b}\mu^-\mu^+$ channel \textcircled{a} 14 TeV with $\cal L$=(100, 600)[$fb^{-1}$]} \\ [0.5ex]
$M_s$[TeV]&\multicolumn{2}{c}{$\delta=2$}&&\multicolumn{2}{c}{$\delta=3$}&&\multicolumn{2}{c}{$\delta=4$}&&\multicolumn{2}{c}{$\delta=5$}&&\multicolumn{2}{c}{$\delta=6$}\\
\hline
3.5   & 60    & 146    && 29  & 70  && 25   &61     && 19       &46   && 15        & 37  \\
4.0   & 26    & 64     && 15  & 36  && 10   &25     && 8        &18.4 && 6         & 14.4  \\
4.5   & 12    & 29     && 7   & 16  && 4.2  &10.3   && 3.2      &8    && 3         & 6  \\
5.0   & 6     & 14     && 4   & 7.3 && 2    &5      && 1.4      &3.4  && 1.1       &3  \\
5.5   & 3     & 7      && 2   & 4   && 1    &2.2    && $\sim 1$ &1.4  && $\sim 1$  &1.2  \\
\hline\hline
\end{tabular}
\end{center}
\vspace*{-0.8cm}
\begin{center}
\begin{minipage}{14cm}
\caption{\label{tab4}  $\frac{S}{\sqrt{B}}$ for the cahnnel $pp\rightarrow HH\rightarrow b\bar{b}\mu^-\mu^+$
after taking into account all backgrounds and adopt the kinematic cuts. }
\end{minipage}
\end{center}
\end{table}

\vskip 5mm
\section{Summary}
\par
In this work, we present the a full treatment of the QCD loop
induced corrections to the 125 GeV neutral Higgs pair production
process $pp\rightarrow ij \rightarrow G_{KK} \rightarrow
HH+X(ij=q\bar{q},gg)$ at the early ($\sqrt{s}=8$ TeV) and future
($\sqrt{s}=14$ TeV) LHC in the context of
large-extra-dimensions(LED) model including the Kaluza-Klein(KK)
excited gravitons. We investigate the dependence of the total
corrected cross sections on the renormalization/factorization scale
$\mu$, and study the corrected distributions of the transverse
momenta($p_T^H$), invariant mass($M_{HH}$) and rapidity($y^{H}$)
distributions. Our final results show that the LED model raises the
cross section of Higgs pair production compare to its SM prediction
and enhance the transverse momentum($p_T^H$) and distributions at
high scales of $p_T^H$ and $M_{HH}$. By including the QCD loop
corrections, the scale dependence of total cross section can be
reduced obviously.  Choose decay modes like $HH\rightarrow b\bar{b}\gamma\gamma$
or $HH\rightarrow b\bar{b}\mu^-\mu^+$ and some simple cuts,
we can strongly reduce the SM background but keep most of the LED effects
leading Higgs pair production a promising channel to search for LED effects.

\vskip 5mm
\par
\noindent{\large\bf Acknowledgments:}
We would like to thank Zhang Ren-You and Li Hong-Lei for useful discussions.
Project supported by the National Natural Science Foundation of
China (Grant No.11147151, No.11205070, No.11105083, No.10947139 and
No.11035003), and by Shandong Province Natural Science Foundation
(No.ZR2012AQ017).


\vskip 10mm
\begin{thebibliography}{99}
\bibitem{ADD}
  N.~Arkani-Hamed, S.~Dimopoulos and G.~R.~Dvali,
  Phys.\ Lett.\ B {\bf 429}, 263 (1998)  [hep-ph/9803315];
  N.~Arkani-Hamed, S.~Dimopoulos and G.~R.~Dvali,
  Phys.\ Rev.\ D {\bf 59}, 086004 (1999)  [hep-ph/9807344].

\bibitem{Kapner:2006si}
  D.~J.~Kapner, T.~S.~Cook, E.~G.~Adelberger, J.~H.~Gundlach, B.~R.~Heckel, C.~D.~Hoyle and H.~E.~Swanson,
  Phys.\ Rev.\ Lett.\  {\bf 98}, 021101 (2007)
  [arXiv:hep-ph/0611184].

\bibitem{HiggsMechanism}
T. Plehn, M. Spira, and P.M. Zerwas, Nucl. Phys. B479, 46 (1996);
Erratum, ibid. B531, 655 (1998).

\bibitem{HH2bbrrbbmm}
U. Baur, T. Plehn and D. L. Rainwater, Phys. Rev. D 68, 033001 (2003);
Phys. Rev. D 69, 053004 (2004).

\bibitem{HH2bbtt}
M. J. Dolan, C. Englert and M. Spannowsky, arXiv:1206.5001 [hep-ph].

\bibitem{sunhaopphh}
Hao Sun, Wen-Gan Ma, Ya-Jin Zhou, Yan-Bin Sun, Ren-You Zhang and
Hong-Sheng Hou, Commun.Theor.Phys.41:73-78(2004).

\bibitem{wangleipphhLH}
Lei Wang, Wenyu Wang, Jin Min Yang, Huanjun Zhang, Phys.Rev.D76,
017702 (2007).

\bibitem{pphhUED}
Hiroshi de Sandes, Rogerio Rosenfeld, Phys.Lett.B659, 323-327
(2008).

\bibitem{pphhED}
C. S. Kim, Kang Young Lee and Jeonghyeon Song,
Phys.Rev.D64:015009(2001).

\bibitem{ADD:Gian}
Gian F. Giudice, Riccardo Rattazzi, James D. Wells, Nucl.Phys. B544,
3-38 (1999).

\bibitem{ADD:HanTao}
Tao Han, Joseph D. Lykken, Ren-Jie Zhang, Phys.Rev.D59:105006(1999).

\bibitem{ADDvirtualA}
J. L. Hewett, Phys. Rev. Lett. 82 (1999) 4765; Prakash Mathews, V.
Ravindran, K. Sridhar and W. L. van Neerven, Nucl. Phys. B713 (2005)
333; Prakash Mathews, V. Ravindran, Nucl. Phys. B753 (2006) 1; M.C.
Kumar, Prakash Mathews, V. Ravindran, Eur. Phys. J. C49 (2007) 599.

\bibitem{ADDvirtualB}
O. J. P. Eboli, Tao Han, M. B. Magro, P. G. Mercadante, Phys. Rev.
D61 (2000) 094007; K.m. Cheung and G. L. Landsberg, Phys. Rev. D 62
(2000) 076003; M.C. Kumar, Prakash Mathews, V. Ravindran, Anurag
Tripathi, Phys. Lett. B672 (2009) 45; Nucl. Phys. B818 (2009) 28.

\bibitem{ADDvirtualC}
  M.~Kober, B.~Koch and M.~Bleicher,
  Phys.\ Rev.\ D {\bf 76}, 125001 (2007)  [arXiv:0708.2368];
  J.~Gao, C.~S.~Li, X.~Gao and J.~J.~Zhang,
  Phys.\ Rev.\ D {\bf 80}, 016008 (2009)  [arXiv:0903.2551];
  Neelima Agarwal, V. Ravindran, V. K. Tiwari, Anurag Tripathi, Nucl.
  Phys. B830 (2010) 248.

\bibitem{ADDvirtualD}
  Z.~U.~Usubov and I.~A.~Minashvili,
  Phys.\ Part.\ Nucl.\ Lett.\  {\bf 3}, 153 (2006)  [Pisma Fiz.\ Elem.\ Chast.\ Atom.\ Yadra {\bf 3}, 24
  (2006)];
  K.~Y.~Lee, H.~S.~Song and J.~-H.~Song,
  Phys.\ Lett.\ B {\bf 464}, 82 (1999)  [hep-ph/9904355].
  Neelima Agarwal, V. Ravindran, V. K. Tiwari, Anurag Tripathi, Phys. Rev. D82 (2010)
  036001;
  B.~Yu-Ming, G.~Lei, L.~Xiao-Zhou, M.~Wen-Gan and Z.~Ren-You,
  Phys.\ Rev.\ D {\bf 85}, 016008 (2012)  [arXiv:1112.4894].

\bibitem{ADDvirtualE}
Prakash Mathews, Sreerup Raychaudhuri, K. Sridhar, Phys. Lett.
B450(1999) 343; JHEP 0007 (2000) 008.

\bibitem{ADDvirtualF}
  K.~Y.~Lee, H.~S.~Song, J.~-H.~Song and C.~Yu,
  Phys.\ Rev.\ D {\bf 60}, 093002 (1999)  [hep-ph/9905227];
  K.~Y.~Lee, S.~C.~Park, H.~S.~Song, J.~-H.~Song and C.~Yu,
  Phys.\ Rev.\ D {\bf 61}, 074005 (2000)  [hep-ph/9910466];
  hep-ph/0105326;
  S.~C.~Inan and A.~A.~Billur,
  Phys.\ Rev.\ D {\bf 84}, 095002 (2011).

\bibitem{pphhEDsunhao}
Hao Sun, Ya-Jin Zhou, He Chen, Eur. Phys. J. C (2012) 72:2011.

\bibitem{CMS:LED1}
  C.~Collaboration [CMS Collaboration],
  arXiv:1204.0821 [hep-ex].

\bibitem{CMS:LED2}
  S.~Chatrchyan {\it et al.}  [CMS Collaboration],
  Phys.\ Lett.\ B {\bf 711}, 15 (2012)  [arXiv:1202.3827 [hep-ex]].

\bibitem{CMS:LED3}
  S.~Chatrchyan {\it et al.}  [CMS Collaboration],
  arXiv:1112.0688 [hep-ex].

\bibitem{LEDbounds}
Roberto Franceschini, Gian Francesco Giudice, Pier Paolo Giardino,
Paolo Lodone, Alessandro Strumia, JHEP 1105, 092 (2011).

\bibitem{SMHiggs125GeV_ATLAS}
ATLAS Collaboration, ATLAS-CONF-2011.163

\bibitem{SMHiggs125GeV_CMS}
CMS Collaboration, CMS PAS HIG-11-032

\bibitem{ADD:ppwwQCD:Mawengan}
Bai Yu-Ming, Guo Lei, Li Xiao-Zhou, Ma Wen-Gan, Zhang Ren-You, Phys.
Rev. D85(2012) 016008.

\bibitem{CT10}
Marco Guzzi, Pavel Nadolsky, Edmond Berger, Hung-Liang Lai, Fredrick
Olness, C.-P. Yuan, SMU-HEP-10-11, [arXiv:1101.0561].

\bibitem{BASES}
S. Kawabata, Comp. Phys. Commun. 88, 309 (1995). F. Yuasa, D.
Perret-Gallix, S. Kawabata, and T. Ishikawa, Nucl. Instrum. Meth.
A389, 77 (1997).

\bibitem{Kaleu}
Hameren A van arXiv:1003.4953[hep-ph].

\bibitem{2PSS:Owens}
B. W. Harris and J. F. Owens,
\newblock Phys. Rev. {\bf D 65}, 094032(2002).

\bibitem{FeynArts}
T. Hahn, Comput.Phys.Commun. 140, 418-431 (2001).

\bibitem{FormCalc}
T. Hahn, Nucl.Phys.Proc.Suppl. 89, 231-236 (2000).

\bibitem{LoopTools}
T.Hahn,M.Perez-Victoria,Comput.Phys.Commun.118:153-165(1999).

\bibitem{Oneloop}
A. van Hameren, Comput.Phys.Commun.182:2427-2438,2011.

\bibitem{HdecayBranch}
A. Djouadi, J. Kalinowski and M. Spira, Comput. Phys. Commun. 108, 56 (1998).

\bibitem{ParticleDataGroup}
Particle Data Group, K. Nakamura et al.,
\newblock JPG {\bf 37}, 075021 (2010).

\bibitem{BaurHHbackground2}
U. Baur, T. Plehn and D. Rainwater, Phys. Rev. D 68, 033001 (2003).

\end{thebibliography}
\end{document}